# The effect of the cosmological expansion on local systems: Post-Newtonian approximation

Jose J. Arenas

*Department of Physics, Monterroso Institute, Santo Tomás de Aquino S/N, 29680, Málaga, Spain*
jjarenasfisica@iesmonterroso.org ; arenasferrer@hotmail.com

**Abstracts**

Frequently, the quantitative effect of the large-scale cosmological expansion on local systems is studied in the light of Newtonian approach, and the General Relativity Theory is neglected. We, however, analyze the influence of cosmological expansion on small systems in the light of Post-Newtonian approximation. The equations show the product $H_0 c = 6.99 x 10^{-10} ms^{-2}$, and it so happens that the magnitude of this acceleration is very close the apparently anomalous acceleration of the Pioneer 10 and 11 spacecraft. Furthermore, we obtain the new radius at which the acceleration due to the cosmological expansion has the same magnitude as the two-body attraction, and the classical critical radius is obtained when the Schwarzschild radius approaches zero (for example, the Solar System).

**I. INTRODUCTION**

The problem of the influence of the cosmic expansion on local gravitationally bound system has been analyzed by several authors [1-9]. The most common approach is to study a Newtonian gravitationally bound system, such as a binary stellar system embedded in an expanding Frieddman-Lemaître-Robertson-Walker (FLRW) universe and computes the effect of the cosmological expansion as a perturbation of the local dynamics ($p_{CE}$).

We consider the dynamical problem of two bodies attracting each other via a force with $1/R^2$ fall-off. For simplicity we may think of one mass as being much smaller than the other one, though this is really inessential. The result for a particle of polar coordinates $(R, \varphi)$ and conserved angular momentum per unit mass $L$ in the field of a mass $M$ embedded in a FLRW universe with scalar factor $a(t)$ is given by the equations of motion:

$$\ddot{R} = -\frac{GM}{R^2} + R\dot{\varphi}^2 + \left(\frac{\ddot{a}}{a}\right) R = -\frac{GM}{R^2} + R\dot{\varphi}^2 + p_{CE}$$
$$R^2 \dot{\varphi} = L, \qquad (1.1)$$

with $G$ the universal gravitational constant.

The global expansion is described by the Hubble law, $\dot{R} = HR$, which states that the relative radial velocity of two comoving objects at a mutual distance $R$ grows proportional to that distance. $H$

denotes the Hubble parameter, it is given in terms of the scalar factor *a(t)*, via $H \equiv \dot{a}/a$. So, the acceleration that results from the Hubble law is given by

$$\dot{v} = \ddot{R} = \dot{H}R + H\dot{R} = \frac{\ddot{a}}{a}R = -qH^2 R$$

(1.2)

where

$$q \equiv -\frac{\ddot{a}a}{\dot{a}^2} = -\frac{\ddot{a}}{a}H^{-2}$$

(1.3)

is the dimensionless deceleration parameter.

Sereno & Jetzer in 2007 [7] and Carrera & Giulini in 2010 [8] analyzed the restricted two-body problem in an expanding universe; deviations from Keplerian orbits in the Solar System, evolution of the orbital radius... In principle, a time varying $\ddot{a}/a$ causes changes in the semi-major axis and eccentricity of Kepler orbits. Others previously [10, 11] considered that on a planet in the Solar System, the relevant time scale of the problem is the period of its orbit around the Sun, and the factor $\ddot{a}/a = -q_0 H_0^2$ is treated as constant during an orbit (the relative error is smaller than $10^{-9}$). Then, the classical critical radius at which the acceleration due to the cosmological expansion has the same magnitude as the two-body attraction is obtained ($R_{CL}$);

$$\frac{GM}{R_{CL}^2} = \left(\frac{\ddot{a}}{a}\right)R_{CL} = AR_{CL} \Rightarrow R_{CL} = \sqrt[3]{\frac{C}{|A|}}$$

(1.4)

with $C = GM$ y $|A| = |q_0|H_0^2$

($H_O \approx 100h Km/s/Mpc$ with $h \approx 0.72$ and $q_0 \approx -1/2$ of the present epoch).

Hence, for $R > R_{CL}$, the effect of the cosmological expansion is the dominant one.

Hence, the effect of the large-scale cosmological expansion on local systems is studied in the light of Newtonian approach, and is neglected the General Relativity (usually). Moreover, in a weak-field approximation ($|V|/c^2 \lesssim 10^{-6}$, with $V$ the newtonian gravitational potential and $c$ the speed of light), the effect the general relativity is computed as a perturbation of the Newtonian mechanics of the Solar System (Post-Newtonian approximation). So, we analyze the influence of cosmological expansion on local systems in the light of Post-Newtonian approximation.

## II. EQUATION OF MOTION

The standard Post-Newtonian equations of motion for a system of $N$ point masses were first obtained by Lorentz and Droste (1917), and they were made famous by Einstein, Infeld, and Hoffmann, who rederived them (1938). Some authors have studied the Solar System by the Post-Newtonian approximation [12-14] and the effect is calculated as a perturbation of the Newtonian dynamics ($p_{PN}$) In the special case in which the system contains only two bodies, the equation of motion reduces to:

$$\ddot{\mathbf{R}} = -\frac{\mu}{R^3}\mathbf{R} + \frac{\mu}{c^2 R^3}\left\{\left[\frac{\mu(4+2\sigma)}{R} - \dot{R}^2(3\sigma+1) + \frac{3\sigma}{2R^2}(\mathbf{R}\cdot\dot{\mathbf{R}})^2\right]\mathbf{R} + (4-2\sigma)(\mathbf{R}\cdot\dot{\mathbf{R}})\dot{\mathbf{R}}\right\} =$$
$$= -\frac{\mu}{R^3}\mathbf{R} + \mathbf{p}_{PN}$$

(2.1)

where $\sigma = \frac{Mm}{(M+m)^2}, \mu = G(M+m)$ and $R \equiv$ mutual distance

If we also assume that $M \gg m$ (test particle model) we obtain: $\sigma \to 0, \mu = GM$, with

$$\ddot{\mathbf{R}} = -\frac{GM}{R^3}\mathbf{R} + \frac{GM}{c^2 R^3}\left\{\left[\frac{4GM}{R} - \dot{R}^2\right]\mathbf{R} + 4(\mathbf{R}\cdot\dot{\mathbf{R}})\dot{\mathbf{R}}\right\} = -\frac{GM}{R^3}\mathbf{R} + \mathbf{p}_{PN}$$

(2.2)

Using the eq. (1.1) and (2.2) (module);

$$\ddot{R} = -\frac{GM}{R^2} + R\dot{\varphi}^2 + p_{CE} + p_{PN}$$

(2.3)

The eq. (2.3) represents the equation of motion for a particle in the field of a mass $M$ embedded in a FLRW universe with scalar factor $a(t)$ in the light of Post-Newtonian approximation.
Furthermore, assuming circular orbits [5, 7, 8], the eq. (2.3) becomes,

$$R\dot{\varphi}^2 = \frac{GM}{R^2} - \left(\frac{\ddot{a}}{a}\right)R - \frac{4G^2M^2}{c^2 R^3}$$

(2.4)

Equating both perturbations for the Solar System (cosmological perturbation and relativistic perturbation);

$$|q_0|H_0^2 R_{CR} = \frac{4G^2M^2}{c^2 R_{CR}^3} \Leftrightarrow R_{CR} = \sqrt{\frac{2GM}{H_0 c\sqrt{|q_0|}}} = 4895 \; AU$$

(2.5)

So, at $R_{CR} = 4895 \; AU$, the cosmological perturbation and the relativistic perturbation are equal. For distances less than $R_{CR}$ (all the planets in the Solar System), the relativistic perturbation is greater than the cosmological perturbation, hence, it should be considered in the influence of the cosmological expansion on local systems.

## III. PIONEER ANOMALY

The analysis of orbital data from Pioneer 10/11 spacecraft indicates the existence of a very weak acceleration ($a_p = [8.74 \pm 1.33] \cdot 10^{-10}\ ms^{-2}$) directed toward the Sun. Since 1998, many explanations have been considered; a consequence of the expanding universe (Rosales & Sánchez-Gómez in 1998 [15] and Lämmerzahl in 2009 [16]), effects such as thermal recoil force (Toth and Turyshev in 2009 [17]), dark matter distribution (Nieto in 2008 [18]), thermal radiation pressure (Rievers & Lämmerzahl in 2011 [19]), a natural consequence of variable speed of light cosmological models (Shojaie in 2012 [20]), anisotropic emission of thermal radiation off the vehicles (Turishev et al. in 2012 [21])…

Moreover, there is an approximate coincidence between the value of this anomalous acceleration and the product:

$$H_0 c = 6.99 \times 10^{-10}\ ms^{-2}$$

(3.1)

That is, the product of the current value of the Hubble constant and the speed of light in vacuum. Note that this product is obtained in (2.5) and it is due to the combination of the cosmological expansion (FLRW universe) and the Post-Newtonian approximation. These Physics frameworks may be related with the anomalous acceleration of spacecraft.

## IV. NEW CRITICAL RADIUS

Now, using (2.3) and (2.4), the new radius ($R_C$) at which the acceleration due to the cosmological expansion has the same magnitude as the two-body attraction is given by the equation:

$$-\frac{GM}{R_C^2} + \left(\frac{\ddot{a}}{a}\right) R_C + \frac{4G^2 M^2}{c^2 R_C^3} = 0$$

(4.1)

Put another way:

$$-GM R_C + |A| R_C^4 + \frac{4G^2 M^2}{c^2} = 0$$

(4.2)

This eq. can be defined in term of parameters of the Schwarzschild radius ($R_\infty = 2GM/c^2$) and the classical critical radius ($R_{CL}^3 = GM/|A|$);

$$R_C^4 + (2R_\infty - R_C) R_{CL}^3 = 0$$

(4.3)

Furthermore, the general solution (real solution with physical meaning) of the eq. (4.3) is given by (Mathematica):

$$R_C = \frac{1}{2}\left[\sqrt{A} + \sqrt{\frac{2R_{CL}^3}{\sqrt{A}} - A}\right]$$

(4.4)

where

$$A = \frac{B}{\sqrt[3]{18}} + \frac{8\sqrt[3]{\frac{2}{3}}R_\infty R_{CL}^3}{B}$$

$$B = \sqrt[3]{\sqrt{3}\sqrt{27R_{CL}^{12} - 2048R_\infty^3 R_{CL}^9 + 9R_{CL}^6}}$$

(4.5)

The solution (4.4) is valid for local systems (see appendix).

Note that if $R_\infty \to 0$ (for example, our Solar System), $R_C = R_{CL}$ is obtained as a particular case of (4.4).

## V. CONCLUSIONS

Frequently, the quantitative effect of the large-scale cosmological expansion on local systems is studied in the light of Newtonian approach, and the General Relativity Theory is neglected. We have obtained the equation of motion for a particle in the field of a mass $M$ embedded in a FLRW universe with scalar factor $a(t)$ in the light of Post-Newtonian approximation. For distances less than $R_{CR}$ (all the planets in the Solar System), the relativistic perturbation is greater than the cosmological perturbation, hence, it should be considered in the influence of the cosmological expansion on local systems. The product $H_0 c = 6.99 x 10^{-10} m s^{-2}$ (Pioneer anomaly) is obtained and it is due to the combination of the cosmological expansion and the Post-Newtonian approximation. It may lead to solving the mystery.

The new critical radius ($R_C$) at which the acceleration due to the cosmological expansion has the same magnitude as the two-body attraction is given by the general solution of equation of motion, and it is valid for local systems.

Hence, the Post-Newtonian approximation should be considered in the influence of the cosmological expansion on local systems.

## VI. ACKNOWLEDGEMENTS

We thank Rosa Sánchez, Miguel García, Edward McCain and Nuria González for their help with the translation of this work.
We are grateful to Eduardo Battaner, Rafael Ortega, and Bert Janssen (University of Granada) for their help to submit this article to arXiv.
We thank Michel Mizony of the Institute Girard Desargues (University of Lyon I) for his comments on classical critical radius.
We also thank Claus Lämmerzahl of the Center Applied Space Technology and Microgravity (ZARM) of the University of Bremen for his references on the Pioneer anomaly.
We also acknowledge the comments on this article about the deceleration parameter made by Domenico Giulini (Institute of Theoretical Physics, Gottfried Wilhelm Leibniz Universität Hannover).

**APPENDIX**

The four solutions are:

1. $R_C = -\frac{1}{2}\left[\sqrt{A} + \sqrt{-\frac{2R_{CL}^3}{\sqrt{A}} - A}\right]$

2. $R_C = \frac{1}{2}\left[-\sqrt{-A} + \sqrt{-\frac{2R_{CL}^3}{\sqrt{A}} - A}\right]$

3. $R_C = -\frac{1}{2}\left[-\sqrt{A} + \sqrt{\frac{2R_{CL}^3}{\sqrt{A}} - A}\right]$

4. $R_C = \frac{1}{2}\left[\sqrt{A} + \sqrt{\frac{2R_{CL}^3}{\sqrt{A}} - A}\right]$

Note that if $R_\infty \to 0$ (for example, our Solar System), we can obtain as a particular case:

$$B = \sqrt[3]{18}\, R_{CL}^2$$
$$A = R_{CL}^2$$

Hence,

1. $R_C = -\frac{1}{2}\left[R_{CL} + \sqrt{-3R_{CL}^2}\right]$
2. $R_C = \frac{1}{2}\left[-\sqrt{-R_{CL}^2} + \sqrt{-3R_{CL}^2}\right]$
3. $R_C = 0$
4. $R_C = R_{CL}$

(4) is the only real solution with physical meaning.